\documentclass[preprint,12pt]{elsarticle}
\usepackage{lineno}
 \usepackage{amsfonts,amssymb,mathtools}
\usepackage{graphicx}
\usepackage[table]{xcolor}
\usepackage{nicematrix}
\DeclareUnicodeCharacter{2212}{-}

\begin{document}

\begin{frontmatter}



\title{Impact on Orbital Period of X-Ray Binary Systems Attached to a Cosmic String}

\author{Ishan Swamy}
\author{Deobrat Singh\corref{cor1}}
\cortext[cor1]{\ead{deobrat.singh@mitwpu.edu.in}}
\affiliation{organization={Department of Physics, Dr. Vishwanath Karad MIT-World Peace University},
            addressline={Kothrud}, 
            city={Pune},
            postcode={411038}, 
            state={Maharashtra},
            country={India}}

\begin{abstract}
Cosmic strings attached to rotating black holes extract its rotational energy, resulting in a mass loss and reduced spin. In this paper we discuss the proposed methods to detect these phenomena and present a novel methodology based on existing literature, by considering a Low Mass X-ray binary system. We investigate the impact of a cosmic string interacting with a black hole in an X-ray binary system and  attempt to explain the observations of unexpected orbital period changes in such systems by proposing mass loss by cosmic strings to be a potential cause. For a period change of order $10^{-10}$, the string tension is $\sim 10^{-17}$, lying in the predicted range for cosmic string tension. An analysis of multiple low mass X-ray binary systems is carried out and it is shown that a significant and observable change occurs for a string tension $\sim 10^{-11}$.
\end{abstract}


\end{frontmatter}


\section{Introduction}

   Cosmic Strings arise from solutions of certain field theories \cite{Vilenkin85} and have also been emerged in String Theory \cite{Polchinski04}. Cosmic Strings are energy densities arising from topological defects resulting in oscillating strings producing stochastic gravitational waves \cite{Vachaspati85}. Detecting these gravitational waves could provide strong evidence of the existence of cosmic strings. A study on the data from the Third Advanced LIGO-VIRGO run has provided an upper limit of the value of string tension, $\mu < 10^{-10}$ \cite{Abbot21}. 

Cosmic strings attached to a spinning black hole extract its rotational energy leading to a reduced spin \cite{Kinoshita16, Igata18}. It is hence suggested that a method to detect cosmic strings is by studying their interaction with black holes. This spin-down phenomenon can be observed by the change in the innermost stable circular orbit (ISCO) of its accretion disk and gravitational waves. Further in another study it has been mentioned that cosmic strings released from galactic nuclei can be detected via gravitational waves, as this phenomenon becomes very observable for a supermassive black hole \cite{Xing2021}. Cosmic strings and primordial black holes have been studied together due to their formation predicted to be in the early universe. Primordial black holes (PBH) attached to cosmic strings can form networks and lead to observable PBH mergers driven by strings \cite{Vilenkin_2018}. 

 An X-ray binary system is a system that consists of a compact star and a companion star that emits X-rays \cite{Thomas06}. The compact star can be either a neutron star or a black hole, whereas the companion star can be either of the following: a main sequence star, a white dwarf, a red giant or supergiant. For convenience, X-ray binary systems will be called X-ray binaries henceforth. X-ray binaries are classified based on the mass of the companion e.g, Low mass X-ray binaries (LMXB) having a companion of mass lower than that of the compact star.

Observations on X-ray binaries discovered a peculiar result - an unexpected change in orbital period. For the binary system Cygnus X-3, the possibility of mass loss due to jets has been discussed to be a cause of this period change \cite{Bhargava17}. However the presence of a third body in the system Cygnus X-3 has also been suggested previously \cite{Igor19}, and a similar result has also been reported for the system LMXB EXO 0748–676 \cite{Corbet94} . 

In this manuscript, we will be discussing the change in the properties of an X-ray binary attached to a cosmic string focusing on its orbital period. We propose here that the extraction of rotational energy by the cosmic string to be one of the potential causes for the unexpected change in the orbital period observed. 

\section{Change in the orbital period due to mass loss}
 In this work we consider a Low Mass X-ray Binary consisting of a spinning black hole and a Main Sequence Star. The string is considered to be in close proximity to the black hole such that the gravitational force of the black hole on the string is far greater than that of the star on the string. Thus, we will not be considering the gravitational interaction between star and cosmic string. We consider a Nambu-Goto cosmic string to be attached to the black hole. Other interactions of Cosmic String with the star will also not be considered.
  \\The attached cosmic string extracts rotational energy from the black hole resulting in a mass loss. As previously discussed, this mass loss impacts the orbital period of the system.
  \\From Kepler's third law,  we have
    \begin{equation}
    P^2 = \frac{4\pi^2b^3}{G_N(M_{BH}+M_*)} \label{Kepler}
    \end{equation}

 where P is the orbital period, $M_{BH}$ is the mass of black hole, $M_*$ is the mass of star, $b$ is the semi-major axis and $G_N$ is the Newton's Gravitational constant,   .
 \\On differentiating (\ref{Kepler}) w.r.t. time, we get
    \begin{equation}
        2P\dot{P} = \frac{4\pi^2}{G_N} \left(\frac{-\dot{m}}{m^2}\right)b^3 + \frac{4\pi^2}{G_N} \left(\frac{3b^2\dot{b}}{m}\right) 
    \end{equation}
    
    \begin{equation}
        2P\dot{P} = \frac{4\pi^2}{G_N} \frac{b^2}{m}\left[3\dot{b}-\frac{b\dot{m}}{m}\right] \\
        = \frac{P^2}{b}\left[3\dot{b}-\frac{b\dot{m}}{m}\right]   
        \label{2}
    \end{equation}
  
    And hence from (\ref{2}) the change in orbital period in terms of $\dot{b}$ and $\dot{m}$ where total mass $m = M_{BH}+M_*$, is
        
    \begin{equation}
        \dot{P} = P\left(\frac{3\dot{b}}{2b} - \frac{\dot{m}}{2m}\right) \label{eq:3}
    \end{equation}
 For an average over one orbit (with almost constant eccentricity), \cite{Hadjidemetriou63, Hadjidemetriou66}
        \begin{equation}
          \frac{\dot{b}}{b} = -\frac{\dot{m}}{m}
        \end{equation}
    This reduces (\ref{eq:3}) to,
    \begin{equation}
            \dot{P} = -2P\frac{\dot{m}}{m} \label{period change}
    \end{equation}
    which is exactly as shown by Simonetti \textit{et al.} \cite{Simonetti_2011}. 
    While rearranging (\ref{Kepler}) we get,
    \begin{equation}
        b = \left(\frac{G_NmP^2}{4\pi^2}\right)^{1/3} \label{semi-major}
    \end{equation} 
    The changes in the semi-major axis w.r.t. time can be calculated as
    \begin{equation}
        \dot{b} = \left(\frac{G_N}{4\pi^2}\right)^{1/3}\left[\frac{1}{3}\left(\frac{P}{m}\right)^{2/3}\dot{m} + \frac{2}{3}\left(\frac{m}{P}\right)^{1/3}\dot{P}\right]\label{change in semi major}
    \end{equation}
Further, the energy extraction rate for a Kerr black hole of mass M with an angular velocity $\Omega_h$ attached to a cosmic string with angular velocity $\omega < \Omega_h$ has been shown \cite{Kinoshita16} as
        \begin{equation}
        \frac{dE}{dt} = 5.8 \times 10^{51} erg/s\left(\frac{G_N\mu/c^2}{1.3 \times 10^{-7}}\right)\left(\frac{q/ac^2}{1/2}\right)\left(\frac{\omega/\Omega_h}{1/2}\right)u(\alpha)
    \end{equation}
     where $a$ is the spin parameter, $q$ is the outward angular momentum flux and $u(\alpha) = 1 - \sqrt{1 - \alpha^2}$ and $\alpha = \frac{ac^2}{G_NM}$ is the dimensionless Kerr parameter.
    Inserting typical values of $a,q,\omega$, the mass loss rate of black hole has been shown \cite{ahmed24} as,
    \begin{equation}
           \dot{m} =  \frac{dM}{dt} \approx -\left(10^4 M_\odot/s\right)\mu\label{10}
    \end{equation}
We notice here that by using (\ref{10}) in (\ref{period change}) a direct relation between period change and cosmic string tension can be established, which had not been attempted previously. Hence the period change due to cosmic string energy extraction is, 
    \begin{equation}
       \dot{P} \approx 2P\left(\frac{10^4 M_\odot/s}{m}\right)\mu \label{12}
     \end{equation}
Interestingly, the period change is directly proportional to cosmic string tension, highlighting the major impact it has on the companion star despite having virtually no interaction with it. The impact of this relation has been explored in the Numerical Analysis section.
We further add here that $\dot{P}$ in (\ref{period change}) does not remain constant, but changes with time $t$ due to mass loss, given by 
    \begin{equation}
        \dot{P}|_t = -2\left(\frac{P\dot{m}}{m + \dot{m}t}\right)
    \end{equation}
It must be noted here that $\dot{P}$ becomes 0 when $t=t_{SD}$ i.e. when the BH's rotational energy has been completely extracted and has been spun down. The orbital period $P$ is at its maximum once this state is reached.
In the case of a supermassive black hole of say $m \approx 10^7 M_\odot$,  $t_{SD} \approx 10^6 yr$ for $\mu = 10^{-11}$ \cite{ahmed24}, and hence $m >> \dot{m}t$ resulting in $\dot{P}$ not being highly observable. However in our case of Low mass X-ray binary, $t_{SD} \approx 1 yr$ making this rate of period change significant in an extremely short timescale. Thus we emphasize here that low mass X-ray binaries are a better system to observe the effects of a cosmic string attached.   

\section{Numerical Analysis}
Observations on certain X-ray binary systems have shown to have orbital period changes which do not have an exact explanation yet. On analysing the observational data for the binary system Cyg X-3 the value of $\dot{P}$ has been estimated \cite{Igor19} as (5.629$\pm$0.002)$\times$10$^{-10}$. If we substitute this value in (\ref{10}) we get the string tension $\mu \sim 10^{-17}$ which lies in the predicted range for cosmic string tension\cite{Turok86}. Further research on other X-ray binaries, such as LMXB EXO 0748–676 \cite{Corbet94, Wolff02} and LMXB AX J1745.6–2901 \cite{Ponti16}, reveals orbital period variations with a $\dot{P}$ of around $10^{-11}ss^{-1}$ which would correspond to $\mu \sim 10^{-18}$. 

We also estimate the period change by considering observational data of certain low mass X-ray binaries by using aforementioned (\ref{12}). Figure \ref{fig:enter-label} gives the relation between the period change, $\dot{P}$ and string tension, $\mu$ for the binary system  4U 1543−47 with black hole of mass $9.4 \pm 2.0 M_\odot$ and companion star of mass $2.45 M_\odot$ orbiting with a period of 26.8 hours \cite{Ritter03, park04, Jonker04J}.\footnote{The plot and data in this work have been generated using Matplotlib library of Python 3 in Jupyter notebook.} The slope remains constant as expected. More systems have been analysed and give the same results for the relation. These results are added in Appendix.

\begin{figure}[th]
    \centering
    \includegraphics[width=0.7\linewidth, height = 7cm]{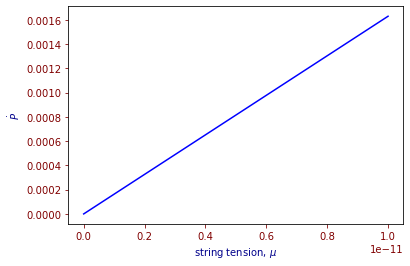}
    \caption{$\dot{P}$ vs $\mu$ for 4U 1543−47}
    \label{fig:1}
\end{figure}
The following Table \ref{Table1} includes observational data for 14 LMXB systems in columns 2, 3, and 4 \cite{Dunn10} and their appropriate references have been mentioned in column 6. Column 5 is our estimations of orbital period change $\dot{P}$ by taking  $\mu = 10^{-11}$, the observable limit for cosmic string tension from gravitational waves \cite{Abbot21} and inserting them in (\ref{12}). This gives us the maximum possible period change values due to the presence of a cosmic string.
Results of Figure:\ref{fig:enter-label} and Table \ref{Table1} have been analysed in the Results and Discussion section.

\begin{table}[pt]
\caption{Analysis of $\dot{P}$ for 14 LMXB}
\label{Table1}
\centering{
\begin{tabular}{||c| c| c| c|>{\color{blue}}c| c||} 

 \hline
\textcolor{red}{Name of} & \textcolor{red}{$M_{BH}$} & \textcolor{red}{$M_*$} & \textcolor{red}{$P$} & \textcolor{red}{$\dot{P}$} & \textcolor{red}{References}\\  
\textcolor{red}{LMXB} & \textcolor{red}{($M_\odot$)} & \textcolor{red}{($M_\odot$)} & \textcolor{red}{$(h)$} & \textcolor{red}{($ss^{-1}$)} & \\[0.5ex]
 \hline\hline
  4U 1543−47 & 9.4 $\pm$ 2.0 & 2.45 & 26.8 & 0.001628 & \cite{Ritter03, park04, Jonker04J} \\ 
 \hline
 4U 1957+115 & [10] & 1.0 & 9.3 & 0.000609 & \cite{Shahbaz96, Thorstensen87}\\
 \hline
 GRO J1655−40 &  7.0 $\pm$ 0.2 & 2.3 & 62.9 & 0.004844 & \cite{Jonker04J, Hynes98a, Shahbaz99} \\
 \hline
 GRS 1758−258 & [10] & [1.5] & 18.5 & 0.001158 & \cite{Smith02}\\
 \hline
 GS 1354−644 & 10.0 $\pm$ 2.0 & 1.02 & 61.1 & 0.003992 & \cite{Ritter03, Casares04}\\ 
 \hline 
 GS 2023+338 & 10 $\pm$ 2 & 0.65 & 155.3 & 0.010499 & \cite{Ritter03, Jonker04J}\\ 
 \hline
 GX 339−4 &  5.8 $\pm$ 0.5 & 0.52 & 42.1 & 0.004796 & \cite{Jonker04J, Hynes03}\\
 \hline
 XTE J1118+480 & 6.8 $\pm$ 0.4 & 0.28 & 4.08 & 0.000415 & \cite{Ritter03, Jonker04J, Wagner_2001}\\
 \hline
 XTE J1550−564 & 10.6 $\pm$ 1.0 & 1.30 & 37.0 & 0.002239 & \cite{Jonker04J, Orosz02}\\
 \hline 
 XTE J1650−500 & 6 $\pm$ 3 & [1.5] & 7.7 & 0.000739 & \cite{Orosz_2004}\\
 \hline
 XTE J1859+226 & 10 $\pm$ 5 & 0.9 & 9.17 & 0.000606 & \cite{Jonker04J, Hynes02}\\
 \hline 
 LMC X1 & 10 $\pm$ 5\ & [1.5] & 93.8& 0.005873 & \cite{Hutchings87, Orosz09}\\ 
 \hline 
 LMC X3 & 10 $\pm$ 5 & 6 & 40.8 & 0.001836 & \cite{Cowley83, Hutchings_2003}\\
 \hline
 SAX 1819.3−2525 & 10 $\pm$ 2 & [1.5] & 67.6 & 0.004232 & \cite{Orosz_2001}\\ [1ex] 
 \hline
\end{tabular}\\
\small
\textit{Note}: Some LMXBs do not have a well-determined mass; hence we have designated $M_{BH} = 10M_\odot$ and $M_* = 1.5M_\odot$ indicated by square brackets.}
\end{table}

\section{Results and Discussion}

In this paper, we described how cosmic strings being attached to a black hole can cause the orbital period of a binary system to change by establishing a direct relation between period change and cosmic string tension. It was emphasized that for a large string tension, the effect is highly observable. We have attempted to explain the observed but unexplained orbital period change in X-ray binaries. 
\\We derived the cosmic string tensions for the observed changes in periods giving a string tension of feasible order $\mu \sim 10^{-17}$. Causes of this period change have been discussed by authors\cite{Corbet94, Wolff02, Ponti16}, but fall short by one or two orders. to explain the period change in the same system for the system LMXB EXO 0748–676, the presence of a third body has been predicted \cite{Corbet94}. Further, the mechanisms of Roche lobe overflow, magnetic braking and mass transfer (accretion) have also been sugested \cite{Wolff02} for the same. Gravitational radiation and magnetic braking have been analysed for the system LMXB AX J1745.6–2901 \cite{Ponti16} but the estimates are one magnitude smaller than the observed value. Among such many explanations we propose here that the mass loss due to cosmic strings could be one of the mechanisms, causing the change in orbital period in X-ray binaries.
\\From Figure:\ref{fig:enter-label} it is clear that there is a significant change in the period when a cosmic string, with sufficiently large $\mu$, gets attached to the system. Further in Table \ref{Table1}, column 5, we notice a period change $\dot{P} \approx 10^{-3} ss^{-1}$ which is a major change when considering they have orbital periods in hours. For example, the period change in the system 4U1543-47 doubles in roughly 2 years. This is a result of the high mass loss rates due to the large string tension and lower mass of the black hole. The spin down time of black holes in LMXBs suggest similar timescales as mentioned in Section 2  \cite{ahmed24}. If such values are observed, it could provide as an indirect evidence for the existence of cosmic strings. These estimates of period change further highlight our point of low mass X-ray binary systems providing for a better detection method. 
\\ It is to be noted here, that the interaction of the companion star and the cosmic string is considered negligible due to the star's low mass. But the same cannot be implemented for a High Mass X-ray binary (for ex. in Cyg-X3) or a black hole binary system, as the companion will have a major gravitational effect on the cosmic string. The system then becomes a three-body problem, which will be explored in our future work. 
\\ Apart from detecting the significant period change, a black hole's change in spin can also be detected by observing its accretion disk formed by the companion star. Black hole spin directly affects the structure and formation of the disk, specifically the radius of innermost stable circular orbit, providing for another detection method for cosmic strings.
\\  There have been many proposed methods to detect cosmic strings, and this paper is an attempt to a better detection method as X-ray binaries are significantly abundant. This method will not only help in detecting cosmic strings but can potentially explain the abnormal period change observed in X-ray binaries.


 \section{Appendix 1}
Following are the plots of the period change vs cosmic string tension for the systems GS 2023+338 ($M_{BH} = 10 \pm 2.0 M_\odot$; $M_* = 0.65 M_\odot$; P = 155.3 hrs) \cite{Ritter03, Jonker04J}, XTE J1118+480  ($M_{BH} = 6.8 \pm 0.4 M_\odot$; $M_* = 0.28 M_\odot$; P = 4.08 hrs) \cite{Ritter03, Jonker04J, Wagner_2001}  and LMC X1 ($M_{BH} = 10 \pm 5 M_\odot$; $M_* = 1.5 M_\odot$; P = 93.8 hrs) \cite{Hutchings87, Orosz09} by using aforementioned (\ref{12}). 
The profile of all the plots is linear, same as Figure \ref{fig:1}.\footnote{The plot and data in this work have been generated using Matplotlib library of Python 3 in Jupyter notebook.}
\begin{figure}[h!]
    \centering
    \includegraphics[width=0.7\linewidth, height = 6cm]{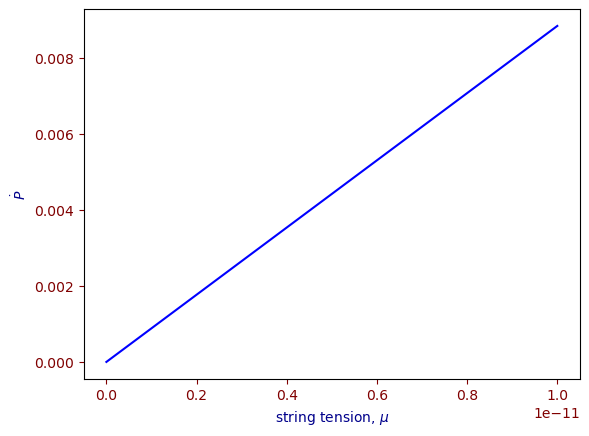}
    \caption{$\dot{P}$ vs $\mu$ for GS 2023+338 }
    \label{fig:enter-label}
\end{figure}
\begin{figure}[h!]
    \centering
    \includegraphics[width=0.7\linewidth, height = 6cm]{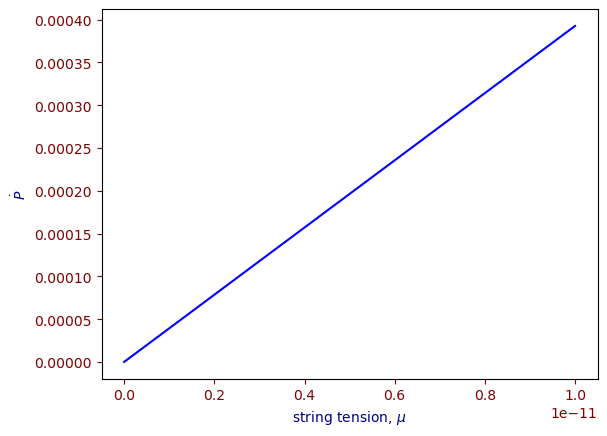}
    \caption{$\dot{P}$ vs $\mu$ for XTE J1118+480}
    \label{fig:enter-label}
\end{figure}
\begin{figure}[h!]
    \centering
    \includegraphics[width=0.7\linewidth, height = 6cm]{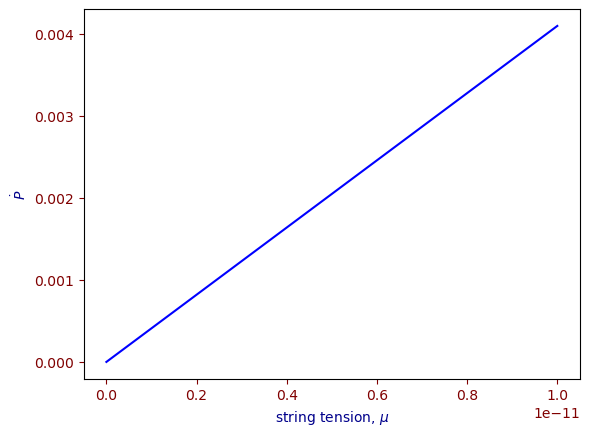}
    \caption{$\dot{P}$ vs $\mu$ for LMC X1}
    \label{fig:enter-label}
\end{figure}

\end{document}